\documentclass[a4paper,10pt]{article}
\usepackage{amssymb}
\usepackage{mathptmx}
\usepackage{graphicx}
\usepackage{amsmath}

\title{A theoretical study of the cluster glass-Kondo-magnetic disordered alloys}
\author{F. M. Zimmer$^1$, S. G. Magalh\~aes$^2$ and B. Coqblin$^3$
\\ \\
{\it $^1$Universidade do Estado de Santa Catarina, 89223-100, Joinville, SC, Brazil}
\\
{\it
$^2$Universidade Federal de Santa Maria, 97105-900, Santa Maria, RS, Brazil}
\\
{\it $^3$ L. P. S., UMR CNRS 8502, Universit\'e Paris-Sud, 91405 Orsay, France.}}

\begin{document}

\maketitle

\begin{abstract}

The physics of disordered alloys, such as typically the well known case of CeNi1-xCux alloys, showing an
interplay  among 
the Kondo effect, the spin glass state and a magnetic order, has been studied firstly
within an average description like in the Sherrington-Kirkpatrick model. Recently, a theoretical model \cite{Magal2}
involving a more local description of the intersite interaction has been proposed to describe the phase
diagram of CeNi1-xCux. This alloy is an example of the complex interplay between Kondo effect and
frustration in which there is in particular the onset of a cluster-glass state. Although 
the model given in Ref. \cite{Magal2} has
reproduced the different phases relatively well, it is not able to describe the cluster-glass state.
We study here the competition between the Kondo effect and a cluster glass phase within a Kondo Lattice
model  with an inter-cluster random Gaussian interaction. The inter-cluster term is treated within the
cluster mean-field theory for spin glasses \cite{Sokoulis}, while, inside the cluster, an exact diagonalisation is
performed including inter-site ferromagnetic and intra-site Kondo interactions. The cluster glass order
parameters and the Kondo correlation function are obtained for different values of the cluster size, the
intra-cluster ferromagnetic coupling and the Kondo intra-site coupling. We obtain, for instance, that the increase of the
Kondo coupling tends to destroy the cluster glass phase. 

\end{abstract}

\section{Introduction}
\label{}
The properties of many cerium or uranium compounds are well described by the Kondo-lattice model,
in which there is a
strong competition between the Kondo effect on each site and the Ruderman-Kittel-Yosida-Kasuya (RKKY) interaction 
between magnetic atoms at different sites. On the other hand, it is now quite clear that the interplay between disorder and 
electronic correlations in these systems produces a new physics such as the presence 
of Non-Fermi liquid behaviour \cite{Miranda1,Miranda2,Castro-Neto} 
or percolative process 
in
the magnetic states similar to manganites \cite{Dagotto}.

An example of the percolative scenario can be found in $Ce$$Ni_{1-x}$$Cu_{x}$ \cite{Marcano1}. This particular alloy presents 
a quite complex phase diagram as long  as the doping of $Ni$ increases. 
In the range of doping 
$0.6<x<0.3$,  the $\mu$SR results indicate formation of clusters  below a 
characteristic temperature $T^{*}$. On the other hand, 
the ac-susceptibility ($\chi_{ac}$)  shows the presence of a glassy ordering below a freezing temperature 
$T_{f}$ ($T_{f}<T^{*}$). In this sense, this particular state can be caracterized as a Cluster Glass (CG) state.  Finally, it has been found an inhomogeneous ferromagnetic (IFM) order from neutron difraction at 
much
lower temperature. 
However, there is no clear indication 
of a Curie temperature $T_{c}$ from $\chi_{ac}$ and $C_{p}$ measurements. 
Therefore, 
it is possible to speculate 
that the evolution from CG state to the IFM order can be obtained by the percolation of the frozen clusters.   One important experimental evidence supporting such scenario has been the behaviour of the hysteresis loops for 
$CeNi_{0.6}Cu_{0.4}$ at $T=100$ mK which  display discrete jumps of the magnetization.

From the theoretical point of view,  
 a model has been proposed 
to explain the presence of frustration in the global phase diagram of $Ce$$Ni_{1-x}$$Cu_{x}$.
The model is a Kondo Lattice with an additional Ising intersite interaction between localized spins called here Kondo-Ising Lattice (KIL) model \cite{Alba1}. The important point is that 
disorder can be
 introduced in the KIL model 
by choosing, 
for instance, the coupling $J_{ij}$ between the localized 
spins 
as a Gaussian random variable
(see for example, the Sherrington-Kirkpatrick (SK) model \cite{SK}).
The results 
have shown that it is possible to 
construct  a mean field solution of the KIL  model, 
where
it is found a  spin glass SG solution as well as a Kondo regime \cite{Alba1}.
This approach has been extended to include a ferromagnetic (FM) solution \cite{Magal1} by displacing the Gaussian distribuition from the origin to $J_{0}$. 
This procedure has
allowed to introduce the usual magnetization as  a new  order parameter. As a result, a global phase diagram temperature {\it versus} 
$J_{K}$ (the  strength of Kondo coupling) has been obtained displaying a SG 
phase,
an additional  FM one and  
a Kondo regime. 
However, the sequence of 
phase transitions when the temperature is decreased (for a constant $J_{K}$) is not in agreement 
to
the experimental findings. On the contrary, the FM solution appears at higher temperature than the SG one.  Moreover, there is no percolation process. The results also show a
conventional phase transitions in which there is, for instance, a clear Curie temperature $T_{c}$. 
It is also important to remark that the usual FM order parameter 
included in 
this particular approach is not able to capture 
the complexity of the experimental IFM ordering.

Recently, a new approach has been proposed 
replacing the random Gaussian $J_{ij}$ in Eq. (\ref{model}) for 
random site model in which  $J_{ij}=\sum_{\mu=1}^{p}\xi^{\mu}_{i}\xi^{\mu}_{j}$,  where $\xi_{i}^{\mu}$  are random variables which follow  the distribution $P(\xi_{i}^{\mu})=1/2\delta_{\xi_{i}^{\mu},1/2}-1/2\delta_{\xi_{i}^{\mu},1/2}$ \cite{Amit1}.
In fact,  this particular choice  of $J_{ij}$ can allow  
the interpolation
from 
weak to strong frustration regimes.
This model improves the previous SK-based model in two directions: first it gives a better possible description of the experimental IFM ordering and second it yelds a disordered ferromagnetic phase below the spin glass one in better agreement with the experiment. However, it is still necessary further theoretical improvements 
to describe the magnetic clusters in 
$Ce$$Ni_{1-x}$$Cu_{x}$. 

The presence of the cluster glass 
is a clear indicative that 
the frustration present in the intermediated doping of $Ce$$Ni_{1-x}$$Cu_{x}$ can not be described by a conventional spin glass.
One possible improvement to the original KIL model 
would be to reformulate the intersite random interaction using  cluster of spins instead of canonical spins. In fact, 
in a earlier work \cite{Sokoulis},
the classical cluster glass problem has been studied 
in a
mean field level.
The model used 
is composed, basically, by a intracluster ferromagnetic coupling 
and
a intercluster Gaussian random coupling. This kind of approach seems 
adequate to be implemented to study the competition between Kondo effect and 
cluster glass
within the approach of Ref. \cite{Alba1}.
Therefore, 
cluster of spins could be introduced 
in the original KIL model \cite{Alba1}
by replacing 
the intersite random Gaussian term 
by 
intracluster and intercluster
terms 
similar to 
Ref. \cite{Sokoulis}. 

The aim of the present work is to study competition between Kondo effect and  cluster glass. 
In view of the discussion in the previous paragraph,
 we 
use
the following 
model 
(called here Kondo Lattice Cluster Glass (KLCG))  to accomplish that study: 
\begin{equation}
\begin{split}
H=&\sum_{a=1}^{N_{cl}}\sum_{i=1}^{n_{s}} \epsilon_{0} \sum_{\sigma=\uparrow\downarrow} \hat{n}_{i\sigma a}^{f} 
- J_{0} \sum_{a=1}^{N_{cl}}\sum_{i<j}^{n_{s}}  \hat{S}_{ia}^{z}\hat{S}_{ja}^{z} + \sum_{a}^{N_{cl}}\sum_{i,j} t_{ij}^{aa} d_{i\sigma a}^{\dagger}d_{j\sigma a} \\+& J_{k} \sum_{a=1}^{N_{cl}}\sum_{i=1}^{n_{s}}\left( \hat{S_{ia}^{+}}s_{ia}^{-} + \hat{S}_{ia}^{-}s_{ia}^{+}\right)- \sum_{a<b}^{N_{cl}} J_{ab} \hat{S}_{a}^{z} \hat{S}_{b}^{z}
\end{split}
\label{model}
\end{equation}
where intercluster coupling $J_{ab}$ is a random variable as the Sherrington-Kirkpatrick model
\begin{equation}
 P(J_{ab})= e^{-J_{ab}^{2}(N_{cl}/32 J^{2})}\sqrt{\frac{N_{cl}}{32\pi J^{2}}}
\label{random}
\end{equation}
with $N=N_{cl}*n_{s}$, where 
$N_{cl}$ and $n_{s}$ are the number of cluster and the number of spin in each cluster, respectively. The hopping 
$t_{ij}^{aa}$
is only inside the cluster.  The indices $(a,b)$  refer  to clusters while $(i,j)$ indicates spins inside a cluster.  So, $S^{+}_{ia}=f^{\dagger}_{ia\uparrow}f_{ia\downarrow}$, 
$s^{-}_{ia}=d^{\dagger}_{ia\downarrow}f_{ia\uparrow}$ and
\begin{equation}
\hat{S}_{a}^{z}=\sum_{i=1}^{n_{s}}\sum_{\sigma=\uparrow\downarrow} \sigma \hat{f}_{i\sigma a}^{\dagger}\hat{f}_{i\sigma a}=\sum_{i=1}^{n_{s}} \hat{S}_{ia}^{z}
\end{equation}
\begin{equation}
\hat{S}_{ia}^{z}=\sum_{\sigma=\uparrow\downarrow} \sigma \hat{f}_{i\sigma a}^{\dagger} \hat{f}_{i\sigma a}
\end{equation}

The problem can be treated within the formalism of integral functional where the spin operators are  given by bilinear combinations of Grassmann fields. 
It should be remarked that disorder is introduced only in the intercluster random Gaussian interaction $J_{ab}$. Nonetheless, 
the thermodynamics for this particular situation  
is also obtained 
using
the replica method.  So, 
\begin{equation}
\beta F=-\frac{1}{N}\lim_{n\longrightarrow 0}\frac{\langle Z^{n} \rangle_{J_{ab}} - 1}{n}~.
\end{equation}

The procedure to deal with the disorder follows closely the the usual fermionic spin glass 
(see for instance \cite{physicaa}). The problem is treated within the static approximation 
with 
order parameters $q_{\alpha\beta}$ ($\alpha\neq\beta$) and $q_{\alpha\alpha}$
being
introduced by a Hubbard-Stratonovich transformation. 
Then,  replica symmetry is assumed $q=q_{\alpha\beta}$ and $p=q_{\alpha\alpha}$. 
The main difference is that, in the present approach, 
these 
order parameters 
describe a glassy ordering among cluster 
instead of
canonical
spins.  
The details of the calculations will be shown elsewhere \cite{clusters}.
\begin{center}
\begin{figure}[th]
\includegraphics[width=13cm,angle=-90]{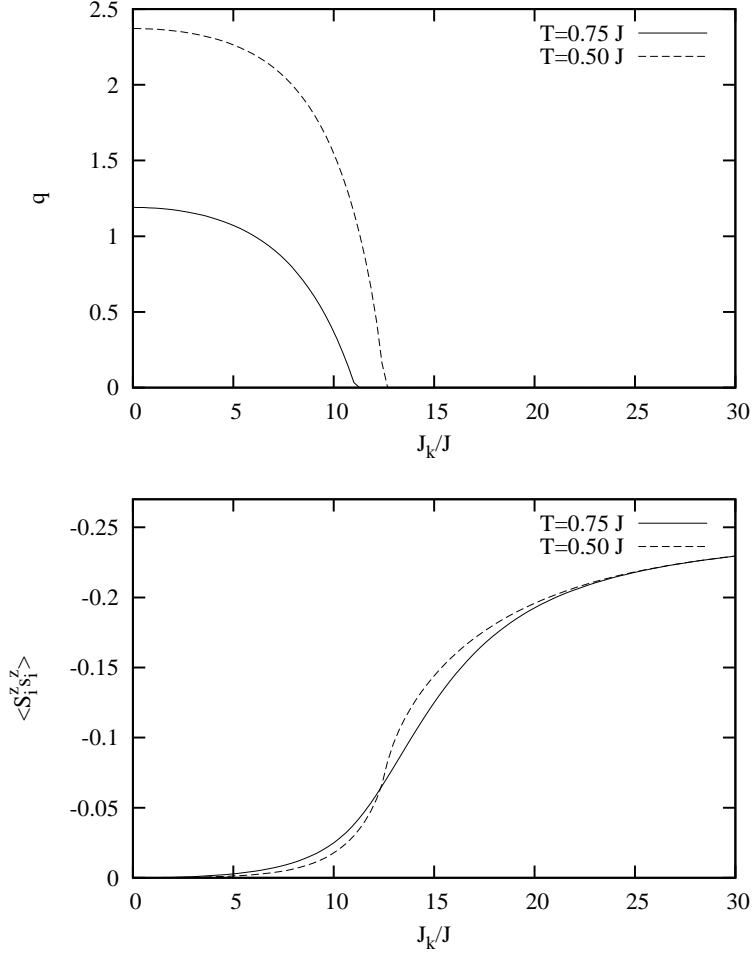}
\caption{Spin glass order parameter and correlaction function $\langle S_i^zs_i^z\rangle$ versus $J_k/J$ for $n_s=2$ and $J_0/J=1$. The dashed lines are results at temperature $T=0.50J$ and the full lines correspond to results at $T=0.75J$.  }
\end{figure}
\end{center}

The free energy is
\begin{equation}
\begin{split}
\beta F =& \frac{(\beta J)^{2}}{2}p^{2} - \frac{(\beta J)^{2}}{2} q^{2}\\ 
-& \int^{+\infty}_{-\infty} \frac{dz}{\sqrt{2\pi}}e^{\frac{-z^{2}}{2}}\ln\left[ \int^{+\infty}_{-\infty} \frac{d\xi}{\sqrt{2\pi}}e^{\frac{-\xi^{2}}{2}}Z_{eff}^{c} \right]
\end{split}
\end{equation}
and
\begin{equation} 
Z_{eff}^{c}=\int D(\varphi^{*}\varphi) D(\psi^{*}\psi)e^{A_{eff}}
\end{equation}

That is equivalent to diagonalize the following Hamiltonian:
\begin{equation}
\begin{split}
H_{eff}=&\sum_{\sigma}\left[\sum_{i=1}^{n_{s}} \epsilon_{0} \hat{n}_{i\sigma} + \sum_{i,j}^{n_{s}} t_{ij} \hat{d}_{i\sigma}^{\dagger}d_{j\sigma} \right] \\ +& J_{k}\sum_{i=1}^{n_{s}}\left[ \hat{S}_{i}^{\dagger} s_{i}^{-} + \hat{S}_{i}^{-} s_{i}^{+} \right] -J_{0} \sum_{i<j}^{n_{s}} \hat{S}_{i}^{z} \hat{S}_{j}^{z} + 2h(p,q)\sum_{j=1}^{n_{s}} \hat{S_{j}^{z}}
\end{split}
\label{Heff}
\end{equation}
where $h(p,q)$ is given as 
\begin{equation}
 h(p,q)=\beta J\sqrt{2(p-q)}\xi + \sqrt{2 q}z.
\end{equation}

Then, 
in 
the effective problem, $h(p,q)$ appears 
as a random external field applied in the cluster which depends on the clusters 
glass order parameters $q$ and $p$. Therefore, 
to obtain 
any information in this problem, it is necessary 
to diagonalize the Hamiltonian $H_{eff}$ given in Eq. (\ref{Heff}) and, simultaneously, to solve the saddle point equations for $q$ and $p$.

\begin{center}
\begin{figure}[th]
\includegraphics[width=12cm,angle=-90]{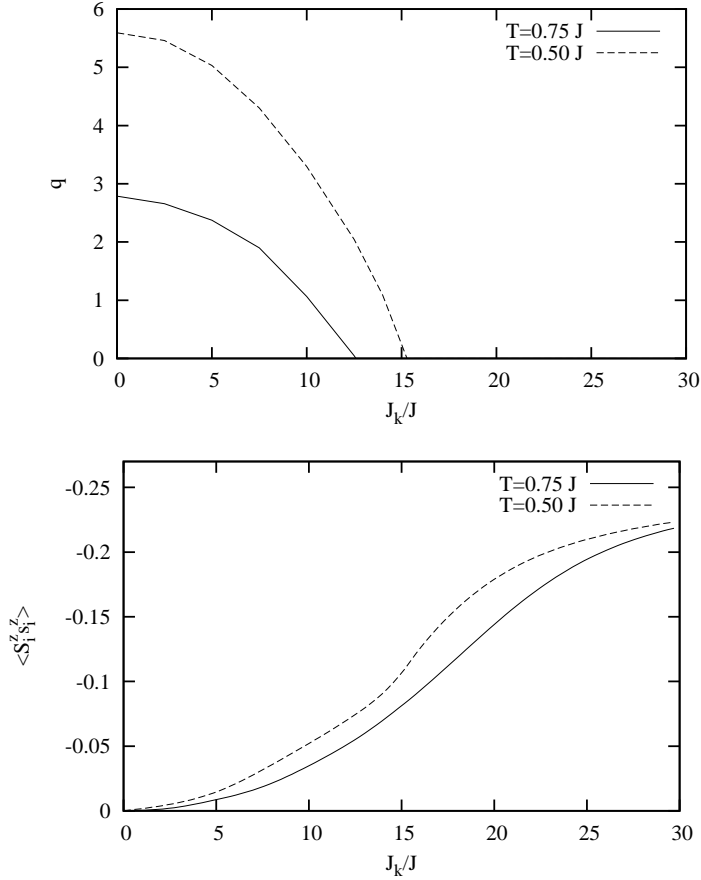}
\caption{Spin glass order parameter and correlaction function $\langle S_i^zs_i^z\rangle$ versus $J_k/J$ for $n_s=3$ and $J_0/J=1$. The dashed lines are results at temperature $T=0.50J$ and the full lines correspond to results at $T=0.75J$.  }
\end{figure}
\end{center}
The behaviour of the 
cluster glass order parameters $q$ and the Kondo correlation function 
$<S^{z}_{i}s^{z}_{i}>$ are displayed in Figs. 1-2 as a function of the Kondo coupling $J_{K}$ 
for two values of temperature $T$ 
while the intracluster ferromagnetic coupling is kept constant $J_{0}=J$ ($J$ is defined in Eq. (\ref{random})).
The size of the clusters also assumes two values, $n_{s}=2$ and $3$.
In Fig. 1, the results for $n_{s}=2$ 
show
 that, 
when $J_{k}$ increases, the cluster glass order parameters $q$ decreases. For $T=0.75J$ and $T=0.5J$, the  
cluster glass phase 
are destroyed 
for $J_{K}\approx 11J$ and $J_{K}\approx 13J$, respectively. 
While $q$ decreases, $<S^{z}_{i}s^{z}_{i}>$ enhances  
 which means that the Kondo effect inside the cluster 
becomes increasingly important. The combined behaviour of $q$ and 
$<S^{z}_{i}s^{z}_{i}>$ would indicate that increase of the Kondo effect inside the cluster 
is
related to the destruction of the cluster glass phase. For $ns=3$, the scenario described previously is
preserved with $q$ decreasing and $<S^{z}_{i}s^{z}_{i}>$ increasing. However, $q$ vanishes for larger values of $J_{K}$ as compared with the case $n_{s}=2$, particularly, for $T=0.5J$. That result would 
suggest that the cluster glass phase becomes more robust with the increase of $n_{s}$

To conclude, in the present work we have studied the competition between cluster glass phase and Kondo effect using the Kondo Lattice Cluster Glass model defined in Eq. (\ref{model}). The intercluster disorder problem is treated using the usual mean field procedure for fermionic spin glasses. Therefore, the original problem is transformed in an effective one in which there is  
 a random external field applied on the cluster.  Finally, it is used exactly 
diagonalization to solve the cluster. The results  indicate that, when $J_{K}$ increases, the Kondo correlation function $<S^{z}_{i}s^{z}_{i}>$ also increases. Simultaneously, the cluster glass phase is destroyed. These results suggest that this 
approach could be used to study the behaviour of the $Ce$$Ni_{1-x}$$Cu_{x}$. However, as discussed previously, the frustration in cited physical system can not be described in terms of a random Gaussian coupling. 
It would be better described by a coupling used in Ref. \cite{Magal2}. This approach is under current investigation.



\end{document}